\documentclass[graybox]{svmult}

\usepackage{type1cm}        
\usepackage{makeidx}         
\usepackage{graphicx}        
\usepackage{multicol}        
\usepackage[bottom]{footmisc}

\usepackage{newtxtext}       %
\usepackage{newtxmath}       

\makeindex             

\begin{document}

\title*{The tachocline revisited}
\author{Pascale Garaud}
\institute{Department of Applied Mathematics, Baskin School of Engineering, UC Santa Cruz, 1156 High St, Santa Cruz, CA 95064, \email{pgraraud@ucsc.edu}}
%
%
\maketitle

\abstract*{The solar tachocline is a shear layer located at the base of the solar convection zone. The horizontal shear in the tachocline is likely turbulent, and it is often assumed that this turbulence would be strongly anisotropic as a result of the local stratification. What role this turbulence plays in the tachocline dynamics, however, remains to be determined. In particular, it is not clear whether it would result in a turbulent eddy diffusivity, or anti-diffusivity, or something else entirely. In this paper, we present the first direct numerical simulations of turbulence in horizontal shear flows at low Prandtl number, in an idealized model that ignores rotation and magnetic fields. We find that several regimes exist, depending on the relative importance of the stratification, viscosity and thermal diffusivity. Our results suggest that the tachocline is in the {\it stratified turbulence} regime, which has very specific properties controlled by a balance between buoyancy, inertia, and thermal diffusion.}

\abstract{The solar tachocline is a shear layer located at the base of the solar convection zone. The horizontal shear in the tachocline is likely turbulent, and it is often assumed that this turbulence would be strongly anisotropic as a result of the local stratification. What role this turbulence plays in the tachocline dynamics, however, remains to be determined. In particular, it is not clear whether it would result in a turbulent eddy diffusivity, or anti-diffusivity, or something else entirely. In this paper, we present the first direct numerical simulations of turbulence in horizontal shear flows at low Prandtl number, in an idealized model that ignores rotation and magnetic fields. We find that several regimes exist, depending on the relative importance of the stratification, viscosity and thermal diffusivity. Our results suggest that the tachocline is in the {\it stratified turbulence} regime, which has very specific properties controlled by a balance between buoyancy, inertia, and thermal diffusion.}

\section{The turbulent tachocline}
\label{sec:1}

Characterizing and understanding the solar tachocline was one of Michael Thompson's many fundamental contributions to the subject of solar physics \cite{ThompsonJCD2003}.
Discovered in the late 1980s \cite{JCDSchou88}, the tachocline is now thought to play a fundamental role in the solar dynamo because it combines substantial 
radial and horizontal shear while being at the interface between the convection zone and the radiation zone. The first model of the solar tachocline \cite{SpiegelZahn1992} was, however, purely hydrodynamical. Noting that the horizontal shear in the tachocline is likely unstable (see \cite{Watson1980,Garaud2001}), Spiegel \& Zahn argued that the latter should be turbulent. In addition, the strong stratification would cause the turbulence to be highly anisotropic, so the transport of angular momentum in the horizontal direction should be much larger than in the vertical direction. With these assumptions, they were able to  propose a simple steady-state model in which the tachocline effectively operates as a turbulent internal boundary layer, across which the latitudinal shear in the convection zone is rapidly reduced down to the negligible levels observed in the radiation zone below. 

To model this idea mathematically, Spiegel \& Zahn argued that the transport of angular momentum in the tachocline should be governed by the following equation, 
\begin{equation}
r^2 \sin^2 \theta \frac{\partial \tilde \Omega}{\partial t} + 2 \Omega_0 r  \cos \theta \sin \theta  u_\theta  \simeq  \frac{\nu_h}{\sin \theta} \frac{\partial }{\partial \theta} \left( \sin^3 \theta \frac{\partial  \tilde \Omega}{\partial \theta}  \right),
\label{eq:SZ1}
 \end{equation}
where vertical advection terms as well as vertical turbulent transport terms have been neglected. In this equation, $r$ and $\theta$ are the radius and co-latitude, respectively, $\Omega_0$ is the mean rotation rate of the Sun while $\tilde \Omega$ is the deviation away from that mean, $u_\theta$ is the latitudinal velocity, and $\nu_h$ is the turbulent horizontal momentum diffusivity. It is important to note that the term on the r.h.s. is a model for the effect of the turbulence -- whether the latter actually behaves in this manner is a question that one should attempt to answer. This equation must be completed with a model for the dynamics of the meridional flow, which can be obtained by considering geostrophic balance and thermal equilibrium: 
\begin{eqnarray}
\frac{1}{\rho} \frac{\partial \tilde{p}}{\partial r} = g \frac{\tilde{T}}{T},  \quad 2 \Omega_0 r \cos \theta \tilde{\Omega} = \frac{1}{\rho r \sin\theta} \frac{\partial \tilde{p}}{\partial \theta}, \\
\mbox{ and } \frac{N^2 T}{ g} u_r = \frac{\kappa_T}{r^2} \frac{ \partial }{\partial r} \left( r^2 \frac{\partial \tilde{T}}{\partial r} \right),
\end{eqnarray}
where $\rho$ and $T$ are the mean background density and temperature in the tachocline, $\tilde p$ and $\tilde T$ are perturbations away from the background means, respectively, $N$ is the buoyancy frequency, $g$ is gravity, and $\kappa_T$ is the local thermal diffusivity. When combined together with incompressibility, these equations ultimately lead to a fourth order differential equation for $\tilde \Omega$, which after some further simplifications (notably, a boundary layer approximation) reads
\begin{equation}
\frac{\partial^4 \tilde \Omega}{\partial r^4} = -  \frac{\nu_h}{\kappa_T} \left(\frac{ N}{2\Omega_0} \right)^2 \left(\frac{\mu}{r_t} \right)^4 \tilde{\Omega} , 
\end{equation}
where $\mu$ is the eigenvalue of an associated latitudinal eigenvalue problem (and is approximately equal to 5), and $r_t$ is the radius of the base of the convection zone. Spiegel \& Zahn showed that the solution of this equation that satisfies appropriate boundary conditions in the radial direction has a characteristic thickness
\begin{equation}
h  = \frac{3\pi}{2\mu} \left(\frac{\Omega_0}{N}\right)^{1/2} \left( \frac{\kappa_T}{\nu_h} \right)^{1/4} r_t \simeq 1.7 \left( \frac{ \nu_h}{\nu} \right)^{-1/4} r_t , 
\end{equation}
using values commonly associated with the tachocline, namely $\Omega_0 \simeq 3 \times 10^{-6}$s$^{-1}$, $N \simeq 10^{-3}$s$^{-1}$, $\nu \simeq 20$cm$^2$/s and $\kappa_T \simeq 2\times 10^7$cm$^2$/s. We therefore see that in order to match observations, where $h$ is estimated to be a few percent of $r_t$, the turbulent viscosity $\nu_h$ in this model needs to be about 8 orders of magnitude larger than the microscopic viscosity $\nu$. A posteriori verification of the conditions under which the model applies also reveals that the ratio of the horizontal to vertical turbulent viscosity should be much larger than $(r_t/h)^2 \sim 10^4$ for the model to be valid, therefore requiring $\nu_r \ll 10^{-4} \nu_h \sim 10^4 \nu \sim 10^5$cm$^2$/s. 

The Spiegel \& Zahn model was later criticized by Gough \& McIntyre \cite{GM98}, who argued by analogy with observations of strongly stratified turbulence in the Earth's atmosphere, that 
\begin{quotation}
 [...] horizontal turbulence controls the distribution of angular momentum in such a way as to drive the system away from, not towards, uniform rotation. Meteorologists once called this negative viscosity. \end{quotation}
In particular, they argued that the use of a turbulent viscosity prescription adopted by Spiegel \& Zahn (leading to equation \ref{eq:SZ1}) is inappropriate, and proposed an alternative model of the tachocline that involves the presence of a large-scale primordial magnetic field embedded in the radiation zone. 
\\
\\
Today, the question of whether turbulent angular momentum transport in the tachocline is diffusive or anti-diffusive remains open. For {\it purely} two-dimensional flows, Tobias, Diamond and Hughes \cite{Tobiasal2007} confirmed that turbulence in the presence of rotation drives large-scale shear rather than quenches it. 
What form it takes in a more realistic three-dimensional system, and whether this conclusion holds or not, remains to be determined. In this paper, I will present recent work on the subject of horizontal shear instabilities in strongly stratified low Prandtl number flows, that will go part-way towards answering these questions.  

\section{Horizontal shear instabilities in stars}
\label{sec:2}

Horizontal shear instabilities in stellar interiors were recently studied by Cope, Garaud and Caulfield \cite{Cope20} in a series of numerical experiments. This section summarizes their results, and presents additional experiments that provide important clues on the nature and relevance of horizontal shear instabilities in stars in general and in the tachocline in particular. 

\subsection{Model description}

In Cope et al. \cite{Cope20}, we consider what is possibly the simplest model for stratified horizontal shear instabilities. We ignore the effect of curvature, rotation, magnetic fields, and compositional stratification, to consider a uniformly thermally stratified, and otherwise triply-periodic Cartesian domain with $z$ pointing in the upward direction. A body force ${\bf F}$ is applied to drive a flow in the streamwise direction $x$, that varies in the spanwise direction $y$, so ${\bf F} = F_0 \sin(ky) {\bf e}_x$. The system of equations governing the flow, in the Spiegel-Veronis-Boussinesq approximation \cite{SV} (which is valid in the tachocline) are:
\begin{eqnarray}
\frac{\partial {\bf u}}{\partial t} + {\bf u} \cdot \nabla {\bf u} = - \rho^{-1} \nabla \tilde p + \alpha \tilde T {\bf g} + \nu \nabla^2 {\bf u} + \rho^{-1} {\bf F}  , \\
\nabla \cdot {\bf u} = 0 ,\\ 
\frac{\partial \tilde T}{\partial t} + {\bf u} \cdot \nabla \tilde T + u_z \left( \frac{\partial T}{\partial z} - \frac{\partial T_{ad}}{\partial z} \right) =  \kappa_T \nabla^2 \tilde T,
\end{eqnarray}
where ${\bf u} = (u_x,u_y,u_z)$, ${\bf g} = - g {\bf e}_z$ is gravity, $\alpha$ is the coefficient of thermal expansion, roughly equal to $1/T$, $\partial T/ \partial z$ is the background mean temperature gradient, and  $\partial T_{ad}/ \partial z = - g/c_p$ is the background adiabatic temperature gradient, where $c_p$ is the specific heat at constant pressure.  The quantities $\tilde T$ and $\tilde p$ are perturbations away from the background state.
 
The equations are non-dimensionalized using the shear lengthscale $k^{-1}$ and the anticipated horizontal velocity of the turbulent flow $U$, obtained by assuming a balance between the forcing and the inertial terms on a scale $k^{-1}$, so 
\begin{equation}
k U^2 = \rho^{-1} F_0 \rightarrow U = \left( \frac{F_0}{k \rho}\right)^{1/2}.
\end{equation}
The temperature scale is taken to be $k^{-1}  \left( \frac{\partial T}{\partial z} - \frac{\partial T_{ad}}{\partial z} \right) $.  The nondimensional equations are then
\begin{eqnarray}
\frac{\partial \hat {\bf u}}{\partial t} + \hat{\bf u} \cdot \nabla \hat{\bf u} = -\nabla \hat p + B \hat T {\bf e}_z + Re^{-1} \nabla^2 \hat {\bf u} + \sin y {\bf e}_x , \label{eq:nondimmom}\\
\nabla \cdot \hat {\bf u} = 0,  \label {eq:nondimdivu} \\ 
\frac{\partial\hat T}{\partial t} + \hat{\bf u} \cdot \nabla \hat T + \hat u_z =  Pe^{-1}  \nabla^2 \hat T, \label{eq:nondimT}
\end{eqnarray}
where 
\begin{equation}
Re = \frac{U}{k\nu}, \quad Pe = \frac{U}{k\kappa_T} = Pr Re, \mbox{ and }  B = \frac{N^2}{k^2 U^2} ,
\label{eq:params}
\end{equation}
are the Reynolds number, P\'eclet number and buoyancy parameters, respectively. Note that in stars, $Pr$ is always much smaller than one (it is for instance of order $10^{-6}$ in the solar tachocilne). In addition, $Re$ is always very large. As a result, we only consider cases where $Pr \ll 1$ and $Re \gg 1$. The remaining parameters, $B$ and $Pe$, can be either large or small. In the  limit of low P\'eclet number, an interesting asymptotic reduction of the equations can be made \cite{Spiegel1962,Lignieres1999}, whereby the dominant balance in the temperature equation is given by 
\begin{equation}
\hat u_z =  Pe^{-1}  \nabla^2 \hat T.
\label{eq:LPNenergy}
\end{equation}
Substituting this into the momentum equation leads to 
\begin{equation}
\frac{\partial \hat {\bf u}}{\partial t} + \hat{\bf u} \cdot \nabla \hat{\bf u} = -\nabla \hat p + BPe \nabla^{-2} \hat u_z {\bf e}_z + Re^{-1} \nabla^2 \hat {\bf u} + \sin y {\bf e}_x,  \label{eq:LPN}\end{equation}
which is called the low P\'eclet number approximation \cite{Lignieres1999}. We see that in that limit, the parameters $B$ and $Pe$ combine to form a single parameter, $BPe$. As such, we expect the flow dynamics to be controlled by the product $BPe$ rather than by $B$ and $Pe$ individually whenever thermal diffusion is important.

\subsection{Low P\'eclet number results}

The study of Cope et al. \cite{Cope20} specifically focusses on the case of small $Pe$, using both the standard governing equations at low $Pe$  (\ref{eq:nondimmom})-(\ref{eq:nondimT}) (here, low $Pe$ is loosely defined as $Pe \le 1$), as well as the low P\'eclet number approximation (\ref{eq:LPN}). The two are found to be in excellent agreement for $Pe \le 0.1$.

\subsubsection{Qualitative results}

\begin{figure}[b]
\sidecaption[t]
\includegraphics[scale=.37]{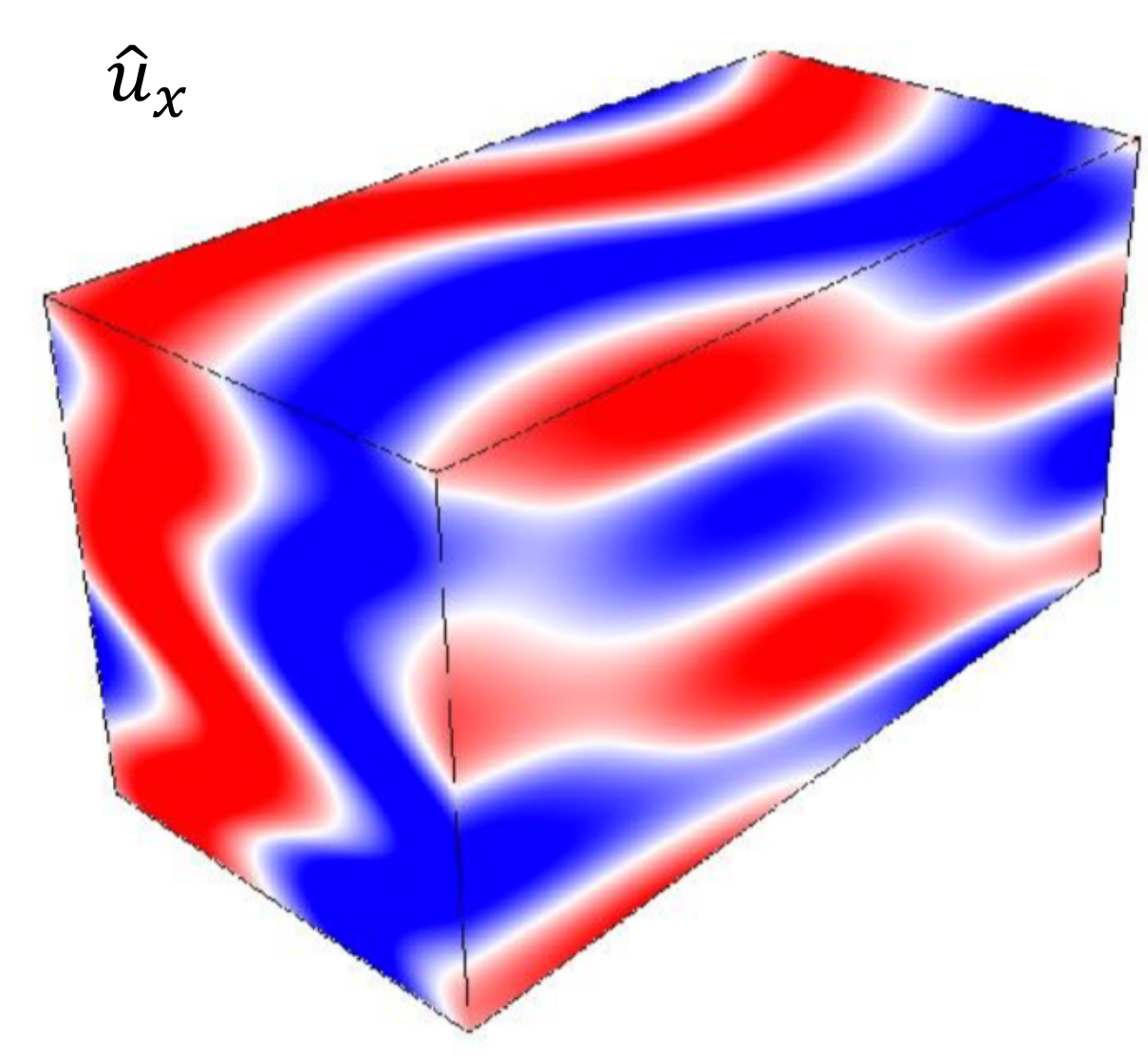}
%
%
\caption{Snapshot of the streamwise flow $\hat u_x$ in a simulation with $Re = 300$, $Pe = 0.1$ and $B = 30,000$, during the initial phase of exponential growth of the instability, showing the presence of horizontal meanders of the basic flow, that are vertically modulated. This creates shear in the vertical direction. Figure adapted from \cite{Cope20}. } 
\label{fig:EarlyTime}       
\end{figure}

In \cite{Cope20}, we present a number of Direct Numerical Simulations (DNSs), evolving the governing equations from some initial conditions until a statistically stationary state is reached (the final state achieved is, to our knowledge, always independent of the initial conditions in this system). In all cases, the domain size is $4\pi \times 2\pi \times 2\pi$. When started from quiescent initial conditions ($\hat {\bf u} = 0$), the imposed force first drives a vertically-invariant flow that is sinusoidal in the $y$ direction. Once it has reached a sufficient amplitude that flow becomes unstable to shear instabilities. The fastest-growing mode is always two-dimensional (2D, i.e. vertically invariant), and is therefore unaffected by the stratification. This 2D mode causes a horizontal meandering of the original shear flow. Depending on the parameter regime, other modes of instability are then also excited, which now depend on $z$. These modes cause a vertical modulation of the phase of the horizontal meanders, and thereby general vertical shear (see figure \ref{fig:EarlyTime}). This vertical shear drives further dynamics that eventually saturate the primary instability. 

\begin{figure}[t]
\sidecaption[t]
\includegraphics[scale=.33]{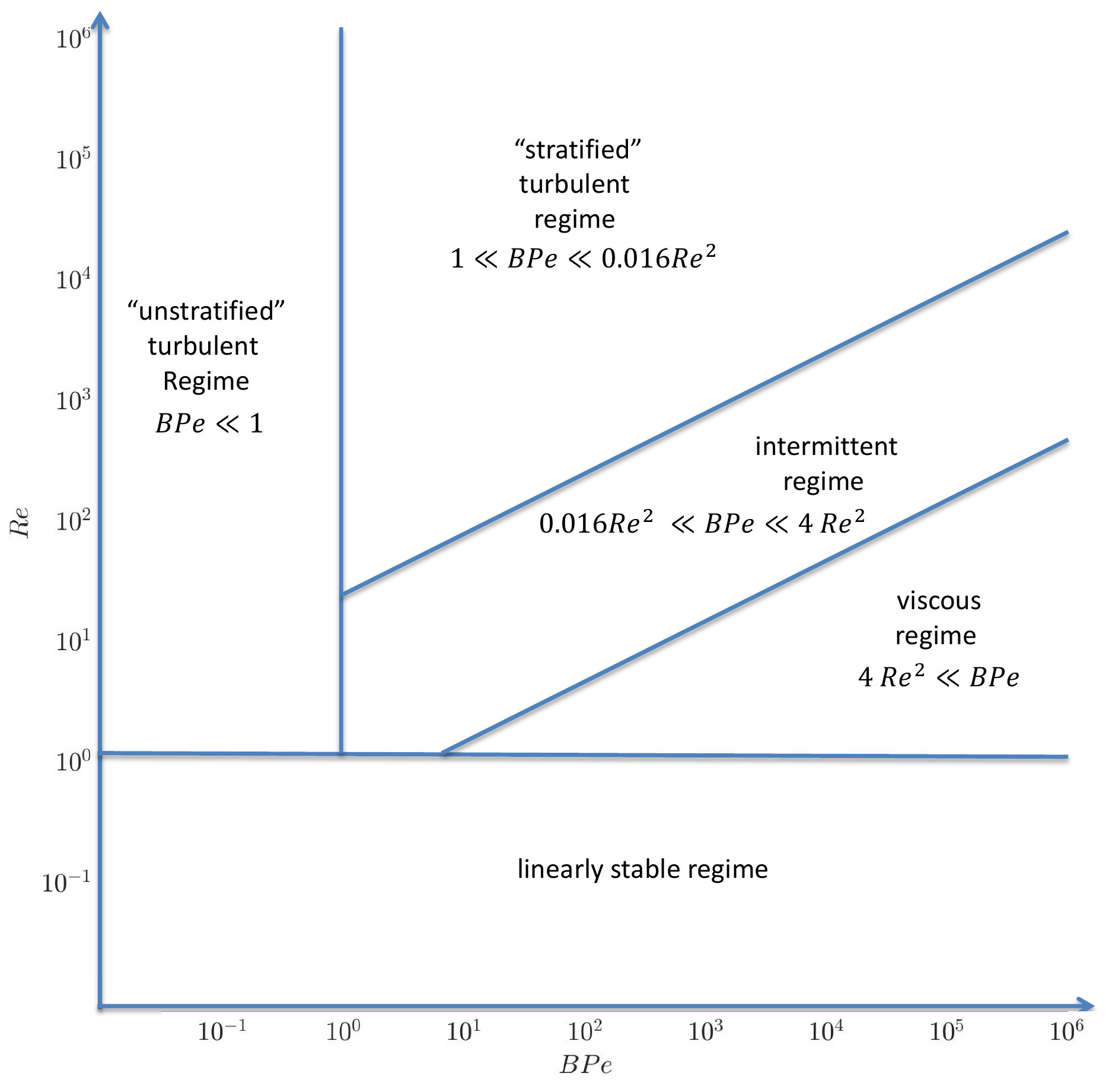}
%
%
\caption{Partitioning of parameter space in the low P\'eclet number limit ($Pe \lesssim 0.1$). Each regime (except the intermittent regime) corresponds to a well-defined dominant balance in the momentum equation, between the forcing, the inertial terms, the buoyancy term, and the viscous terms. The intermittent regime contains regions that are viscously-dominated as well as regions that are in the stratified turbulent regime. Figure from \cite{Cope20}. }
\label{fig:parameterspace}       
\end{figure}

\begin{figure}[b]
\includegraphics[scale=.4]{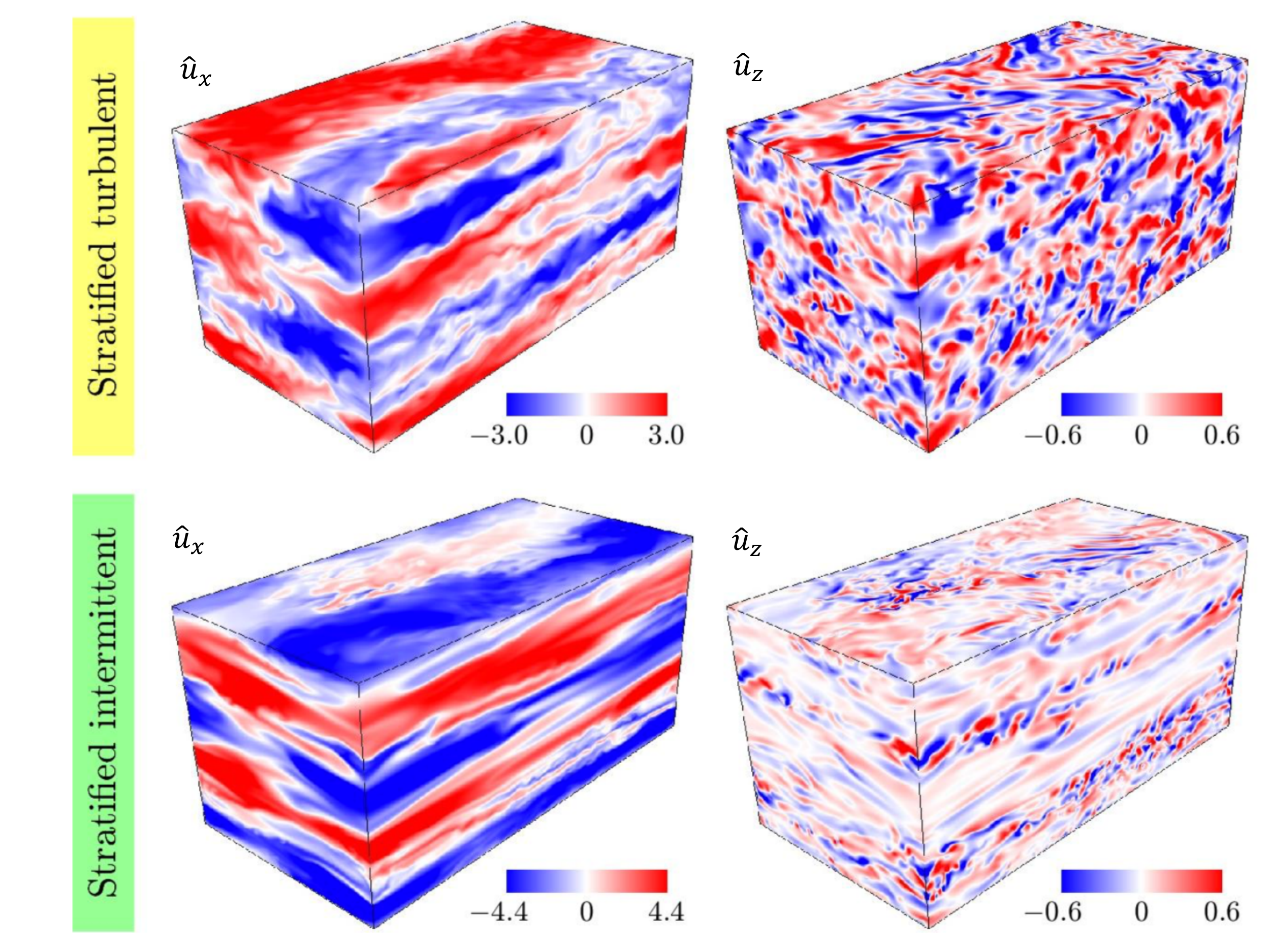}
%
%
\caption{Snapshots of $\hat u_x$ and $\hat u_z$ in the stratified turbulent regime (top, $B = 100$) and intermittent regime (bottom, $B = 10,000$) at $Re = 300$, $Pe = 0.1$. Note how the meanders of the streamwise flow are still visible even in the fully turbulent flow, and note the small vertical scale of the turbulent eddies in both cases.
Figure adapted from \cite{Cope20}. }
\label{fig:cases}       
\end{figure}

We found that the system dynamics, once a statistically stationary state is achieved, only depend on $Re$ and the product $BPe$, as expected from the argument above. We also found that there are (at least) 4 distinct regimes (in addition to the laminar solution at low Reynolds number) depending on the respective values of $Re$ and $BPe$, see figure \ref{fig:parameterspace}. The boundaries between the various regimes can be determined from dominant balance arguments when possible, and empirically otherwise (see \cite{Cope20} for more detail).
Snapshots of two simulations taken in the stratified turbulence regime and stratified intermittent regime are shown in figure \ref{fig:cases}. 
  In the stratified turbulent regime, vertical shear instabilities ubiquitously develop between the meanders of the streamwise flow. The vertical size of the eddies is controlled by a balance between buoyancy, inertia, and thermal diffusion, see Section \ref{sec:LPNscaling}. Interestingly, we find that the eddies are relatively isotropic, with a horizontal scale that is commensurate with the vertical scale.

  As stratification continues to increase, the system enters a regime where turbulence is only found intermittently (both in time and space). Turbulence within these localized patches is similar that found in the stratified turbulence regime. On the other hand, the effect of viscosity becomes important outside of these patches, which are increasingly sparse as $BPe$ increases. For sufficiently large $BPe$, the system enters a viscously dominated regime.


\subsubsection{Scaling laws}
\label{sec:LPNscaling}

\begin{figure}[b]
\includegraphics[scale=.5]{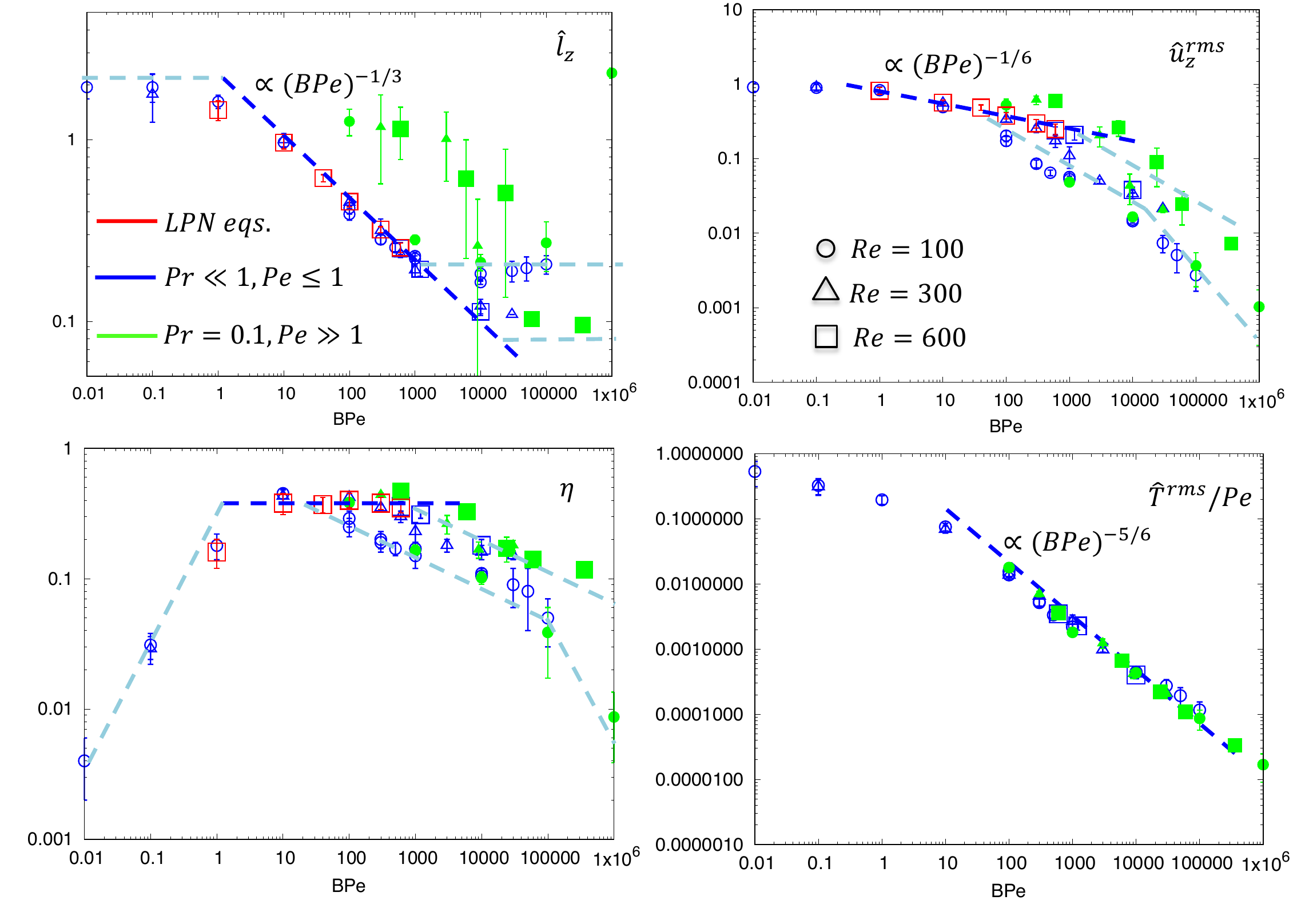}
%
%
\caption{Diagnostic quantities extracted from all available simulations. The open blue and red symbols correspond to simulations at very low Prandtl number and low $Pe$ ($Pe \le 1$), see figure for legend. The green filled symbols correspond to simulations with $Pr = 0.1$, high $Pe$. Errorbars are added in all cases, but are sometimes smaller than the symbol. The dashed lines correspond to the model scaling laws. Emphasis is put on the stratified turbulence regime (dark dashed lines), see \cite{Cope20} for information on the other regimes (pale dashed lines). }
\label{fig:data}       
\end{figure}

In order to gain a more quantitative understanding of the dynamics of the various regimes identified in the low P\'eclet number case, we extracted various diagnostics from the DNSs. In particular, we measured the vertical eddy scale $\hat l_z$ (using the autocorrelation function of the vertical velocity field), the r.m.s. vertical velocity $\hat u_z^{\rm rms} = \langle \hat u_z^2 \rangle^{1/2}$, the r.m.s. temperature $\hat T^{\rm rms}$ and the mixing efficiency $\eta$, which is defined as
\begin{equation} 
\eta = \frac{  -B \langle \hat u_z \hat T \rangle }{-B \langle \hat u_z \hat T \rangle + Re^{-1} \langle | \nabla \hat {\bf u}|^2 \rangle} ,
\end{equation}
where $\langle \cdot \rangle$ denotes an average over the entire computational domain. These quantities were then time-averaged during the statistically stationary phase, and the results are shown in figure \ref{fig:data}. The significance of $\eta$ can be understood by noting that in a statistically stationary state, the kinetic energy equation (which is obtained by dotting the momentum equation by $\hat {\bf u}$ and integrating over the domain, using periodicity to eliminate boundary terms) reduces to 
\begin{equation}
\langle \hat {\bf u} \cdot \hat {\bf F} \rangle  = - B \langle \hat u_z \hat T \rangle + Re^{-1} \langle | \nabla \hat {\bf u}|^2 \rangle .
\end{equation}
We therefore see that the energy input into the system by the force $\hat {\bf F}$ (on the l.h.s.) is either converted into potential energy (first term on the r.h.s.) or viscously dissipated (second term on the r.h.s.).  As such, the ratio $\eta$ measures how efficiently the total energy input into the system is used to mix the background stratification. This quantity turns out to be an excellent diagnostic of the properties of the flow, as shown by \cite{Cope20}.  

In the stratified turbulent regime, which is of potential relevance to stellar interiors,  we found that the vertical lengthscale of the turbulent eddies scales 
as $\hat l_z \sim (BPe)^{-1/3}$. This scaling can be understood from a simple dominant balance in the vertical component of the momentum equation between the nonlinear terms and buoyancy term, $\hat {\bf u} \cdot \nabla \hat u_z \simeq B\hat T$. Meanwhile from the low P\'eclet number thermal energy equation, we have $\hat u_z \simeq Pe^{-1} \nabla^2 \hat T$. 
From a scaling perspective, these two equations result in
\begin{equation}
\frac{\hat u^{\rm rms}_h \hat u^{\rm rms}_z}{\hat l_z} \sim B \hat T^{\rm rms}  \mbox{   and  } \hat u^{\rm rms}_z \sim Pe^{-1} \frac{ \hat T^{\rm rms}}{\hat l_z^2} ,
\label{eq:LPNscaling1}
\end{equation}
where $\hat u^{\rm rms}_h  \sim O(1)$ is the expected horizontal velocity in the non-dimensionalization chosen. Combined, this results in $\hat l_z \sim (BPe)^{-1/3}$, as observed. 

In the same regime, we also observe that $\eta$ is constant and roughly equal to 0.4, which shows that about 40\% of the total energy input into the system is spent mixing the background stratification, while about 60\% is dissipated viscously. By construction, the non-dimensional energy input rate is of order one, so we find that $B \langle \hat u_z \hat T \rangle = O(1)$ as well, which implies that $B \hat u^{\rm rms}_z \hat T^{\rm rms} = O(1)$. Combining this with (\ref{eq:LPNscaling1}) above, we then predict that 
\begin{equation}
 \hat u^{\rm rms}_z \sim (BPe)^{-1/6} \mbox{   and  }  \hat T^{\rm rms} \sim Pe (BPe)^{-5/6}.
\label{eq:LPNscaling2}
\end{equation}
Both scalings are indeed observed in high Reynolds number and low P\'eclet number simulations (see figure \ref{fig:data}).  The prefactor in both cases is found to be of order unity.

As discussed by Cope et al. \cite{Cope20}, the stratified turbulent regime is valid for  $Pe \lesssim 0.1$ and $1 \lesssim BPe \lesssim 0.0016Re^2$. When $BPe$ increases beyond that threshold, the fraction of the domain that is turbulent decreases and is gradually replaced by laminar regions that are viscously dominated. 

\subsubsection{Mixing by horizontal shear flows at low P\'eclet number}

If a star is known to exhibit a horizontal shear flow with amplitude $U$ and wavenumber $k$, then one can easily construct the parameters $Re$, $Pe$ and $B$ from (\ref{eq:params}). If the P\'eclet number is then found to be small, the scalings obtained above can be expressed dimensionally to yield
\begin{eqnarray}
&& l_z \sim (BPe)^{-1/3} k^{-1} \sim \left(\frac{N^2}{U \kappa_T}\right)^{-1/3},   \\
&& u^{\rm rms}_z \sim  (BPe)^{-1/6}  U  \sim   \left(\frac{N^2}{U^7 k^3 \kappa_T}\right)^{-1/6},
\label{eq:scalings1}
\end{eqnarray}
with proportionality constants of order unity. A vertical mixing coefficient can then be formed as
\begin{equation}
D \sim  l_z  u^{\rm rms}_z \sim  \left(\frac{N^2}{U k^3 \kappa_T}\right)^{-1/2} Uk^{-1} .
\label{eq:scalings2}
\end{equation}

\subsection{Low Prandtl number / high P\'eclet number results}

Unfortunately, the results described in the previous section are not  a priori applicable to the solar tachocline because the latter is not a low P\'eclet number shear layer, at least when the P\'eclet number is computed using properties of the mean flow. Indeed, using a typical lengthscale $k^{-1} \simeq r_t/4$ and typical velocity $U = r_t \Delta \Omega$, where $r_t \simeq 5 \times 10^{10}$cm and $\Delta \Omega \simeq 3\times 10^{-7}$s$^{-1}$ is the difference between the angular velocity of the equator and the pole, we find that 
\begin{equation}
Re \sim O(10^{13}),  \quad Pe \sim O(10^7),  \mbox{   and   } B \sim O(10^6),
\end{equation}
so clearly $Pe \gg 1$. In that limit, the low P\'eclet number approximation that is central to the derivation of the scaling laws presented in the previous section does not apply {\it a priori}.
To see how having a large P\'eclet number modifies the results, we ran a number of simulations at $Pr = 0.1$ and large Reynolds number, so that $Pe = 0.1Re$ remains large. 
This is computationally challenging, since a high P\'eclet number with a low Prandtl number requires an even larger Reynolds number, which demands very high resolution. The preliminary results presented here are therefore limited to $Re = 100 (Pe = 10)$, $Re = 300 (Pe=30)$ and $Re = 600 (Pe = 60)$.

\begin{figure}
\includegraphics[scale=.37]{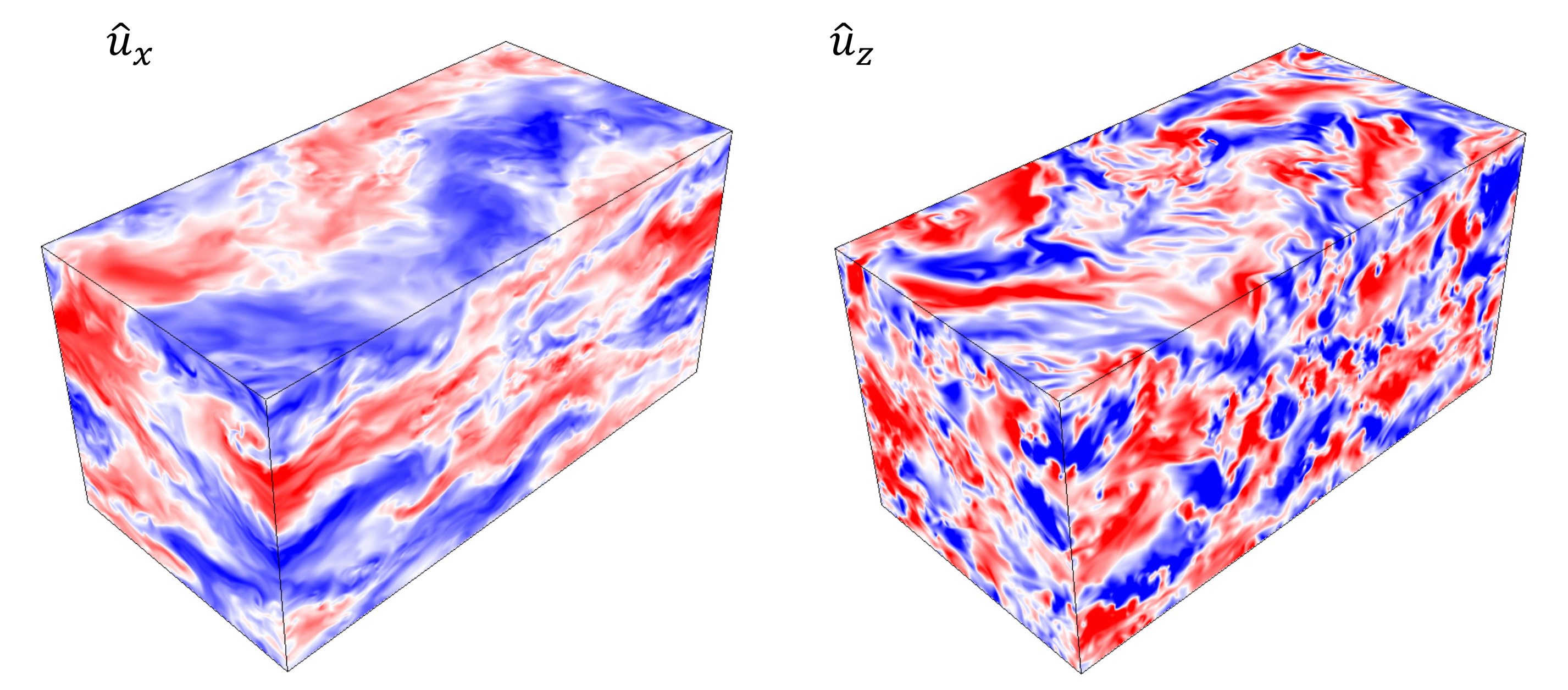}
%
%
\caption{Snapshots of the horizontal velocity $\hat u_x$ and vertical velocity $\hat u_z$ taken in the statistically stationary state of a simulation with $Re = 600$, $Pe = 60$ and $B = 10$, which is in the stratified turbulence regime.}
\label{fig:hipe}       
\end{figure}

Surprisingly, we found that many of the trends observed in low P\'eclet number flows continue to hold in this case, at least qualitatively speaking. In particular, we find that the instability of the base flow continues to give rise to vertically-modulated meandering of the streamwise velocity field, that then creates strong vertical shear. For weak to moderate stratification, the vertical shear becomes in turn unstable, causing vertical mixing. The same regimes can be identified (unstratified turbulence, stratified turbulence, intermittent and viscous). In the stratified turbulent regime, snapshots of the flow look qualitatively very similar to those obtained in the low P\'eclet number regime (see figure \ref{fig:hipe}). 

  In each case, we extracted again quantities such as $\hat l_z$, $\hat T^{\rm rms}$, $\hat u_z^{\rm rms}$ and $\eta$. Results for moderate to large $B$ are presented in figure \ref{fig:data} (green symbols) and  reveal a number of very interesting findings. Crucially, we find for instance that the data for $\hat T^{\rm rms} $ collapses onto the same approximate scaling of  $\hat T^{\rm rms} \sim Pe (BPe)^{-5/6}$ found in the low P\'eclet limit, suggesting that the flow dynamics continue to be thermally diffusive even though $Pe$ is large.  We also find that the new high P\'eclet number data collapse with the low P\'eclet number data in the viscous regime (this is particularly apparent for the $Re = 100$ runs (circles), for which the viscous regime starts around $BPe \simeq 1000$). Finally, and most importantly, our results tentatively suggest that the {\it same} scaling laws apply in the stratified turbulent regime, albeit with a different pre-factor that depends on the Prandtl number. 
  
  These results are surprising at first, since the scaling laws derived in Section \ref{sec:LPNscaling} require the low P\'eclet number approximation to the thermal energy equation (\ref{eq:LPNenergy}) to hold, which should not be the case at high $Pe$. 
  To understand why these low P\'eclet number scalings might still apply to the low Prandtl number / high P\'eclet number cases, it is important to remember that the derivation of the low P\'eclet approximation \cite{Lignieres1999} relies on the assumption that the {\it turbulent} P\'eclet number $Pe_l  = U l / \kappa_T$ be small, where $Pe_l$ is computed using the r.m.s. velocity of the fluid (for which $U$ is a good approximation, at least in the horizontal direction) and the {\it actual} eddy scale $l$. By contrast, the input parameter $Pe$ defined in (\ref{eq:params}) is based on the largest possible physical scale of the system, which is that of the imposed shear, $k^{-1}$. We have seen that the emergent scale of the eddies in the stratified turbulent regime is much smaller than $k^{-1}$ in all directions, so it is quite likely that our simulations are in a regime where $Pe_l \lesssim 1$. This would explain why the emergent dynamics remain thermally diffusive even though $Pe \gg 1$. 

Based on the very limited simulations available at high P\'eclet number / low Prandlt number, we therefore argue that the scalings in the stratified turbulent regime likely remain the same as those described and derived in Section \ref{sec:LPNscaling}, with the exception of constant prefactors that depend weakly on $Pr$ (perhaps logarithmically so), and tend to the ones obtained in the low P\'eclet limit when $Pr \rightarrow 0$. The regime boundaries for the stratified turbulent regime remain to be determined, but it is quite likely that these will also only depend logarithmically on $Pr$, when $Pr \ll 1$. To test these predictions will require simulations at lower Prandtl number and ideally higher P\'eclet number, which will be very challenging computationally, but not impossible. If they are confirmed, this  will have important consequences for stellar interiors.

\section{Discussion}

The findings presented in this paper suggest that horizontal shear instabilities in stars (i.e. at low Prandtl number) generate vertically modulated meanders of the basic flow, on vertical scales that are sufficiently thin to be thermally diffusive. As a result, diffusive vertical shear instabilities can develop, and give rise to small-scale turbulence and vertical mixing. The turbulence is relatively isotropic on the small scales, but becomes more anisotropic on the larger horizontal scales associated with the meanders, as illustrated in figures \ref{fig:cases} and \ref{fig:hipe}. 

This is quite different from the effect of horizontal shear flows in geophysical systems, where the Prandtl number is large. Indeed, in that case thermal diffusion is always negligible if the flow is turbulent (since $\kappa_T \lesssim \nu$). This implies that secondary vertical shear instabilities cannot be excited if the stratification is strong, but are instead limited to localized regions where the stratification is weakened. By and large, the turbulence therefore remains almost two-dimensional, and in a rotating system, this two-dimensional turbulence would indeed have an anti-diffusive behavior, as seen in the Earth's atmosphere and invoked by Gough \& McIntyre to argue against the Spiegel \& Zahn model of the tachocline. 
Our simulations at low Prandtl number however demonstrate that the turbulence is clearly three-dimensional  and almost isotropic on the small scales, and only become anisotropic on the larger scales. Whether this ultimately behaves in a diffusive or anti-diffusive manner in the presence of rotation therefore remains to be determined. 

Tentatively, we propose new scaling laws for the vertical eddy scale, vertical velocity, and vertical mixing coefficient in horizontal shear instabilities in the stratified turbulent regime, given in equations (\ref{eq:scalings1}) and (\ref{eq:scalings2}). The prefactors are of order unity when the P\'eclet number is small (see \cite{Cope20} for more detail), but might depend logarithmically on $Pr$ if $Pr \ll 1$ but $Pe \gg 1$. If this is confirmed then simple order-of-magnitude estimates for $l_z$, $u_z^{\rm rms}$, $T^{\rm rms}$  and $D$ in the tachocline are
\begin{equation}
l_z \sim O(10 {\rm km}),  u_z^{\rm rms} \sim O(10 {\rm cm/s}), T^{\rm rms} \sim O(100{\rm K}) \mbox{  and } 
D \sim O(10^7 {\rm cm}^2/{\rm s}),
\end{equation}
since the tachocline is likely in the stratified turbulent regime (with $BPe \sim 10^{13} \ll 0.002 Re^2$). With this estimate, we see that one of the fundamental assumptions of the Spiegel \& Zahn model may not be satisfied: indeed, if the vertical turbulent momentum diffusivity is of order $
\nu_r \sim D \sim 10^7$cm$^2$/s, then it is too large to neglect in equation (\ref{eq:SZ1}), and would therefore invalidate the model. A possible solution to the problem (other than Gough \& McIntyre's magnetic idea) is that the model holds with $\nu_h \gg 10^8 \nu$, and that the actual tachocline is in fact much thinner than $\sim 0.01 r_t$ -- this cannot be ruled out by observations. 

Of course, much remains to be done to characterize mixing by shear instabilities in the tachocline (and in stars more generally). Several effects have indeed been neglected, that will have to be included before definite conclusions can be made. First, it will be important to include the effect of rotation. Indeed, while the latter is not expected to influence the turbulence on the smaller scales (where the Rossby number is large), it could alter or even suppress the development of the primary instability, which occurs on the larger scales. Second, magnetic fields will also need to be included, since they are expected to be present and significant in the tachocline. Finally, the tachocline has both large-scale vertical and horizontal shear, and the former may influence the nonlinear saturation of the latter. 

\begin{acknowledgement}
This work was initiated as a project at the Woods Hole Geophysical Fluid Dynamics summer program in 2018; we thank the GFD program for their support. P.G. is funded by the National Science Foundation (NSF AST 1814327), and wishes to thank L. Cope, C.P. Caulfield and D. Gough for stimulating discussions.
\end{acknowledgement}

\bibliography{biblio}

\end{document}